\documentstyle[12pt]{article}
\begin{document}
\title
{ Screening the graviton background, graviton pairing, and
Newtonian gravity}
\author
{By Michael A. Ivanov \\
Dept. of Physics, \\
Belarus State University of Informatics and Radioelectronics, \\
6 P. Brovka Street,  BY 220027, Minsk, Republic of Belarus.\\
E-mail: ivanovma@gw.bsuir.unibel.by.}

\maketitle

\begin{abstract}
It is shown that screening the background of super-strong
interacting gravitons creates for any pair of bodies as an
attraction force as well an repulsion force due to pressure of
gravitons. For single gravitons, these forces are approximately
balanced, but each of them is much bigger than a force of
Newtonian attraction. If single gravitons are pairing, a body
attraction force due to pressure of such graviton pairs is twice
exceeding a corresponding repulsion force under the condition that
graviton pairs are destructed by collisions with a body. If the
considered quantum mechanism of classical gravity is realized in
the nature, then an existence of black holes contradicts to
Einstein's equivalence principle. In such the model, Newton's
constant is proportional to $H^{2}/T^{4},$ where $H$ is the Hubble
constant, $T$ is an equivalent temperature of the graviton
background. The estimate of the Hubble constant is obtained
$H=2.14 \cdot 10^{-18} \  s^{-1}$ (or $66.875 \  km \cdot s^{-1}
\cdot Mpc^{-1}$).
\end{abstract}

\section[1]{Introduction }
It was shown by the author in the previous study \cite{1,2} that
an alternative explanation of cosmological redshift as a result of
interaction of a photon with the graviton background is possible.
In the case, observed dimming of supernovae Ia \cite{3} and the
Pioneer 10 anomaly \cite{4} may be explained from one point of
view as additional manifestations of interaction with the graviton
background. Some primary features of a new cosmological model,
based on this approach, are described in author's preprint
\cite{4a}.
\par
In this paper, forces of gravitonic radiation pressure are
considered which act on bodies in a presence of such the
background. It is shown that pressure of single gravitons of the
background, which run against a body pair from infinity, results
in mutual attraction of bodies with a magnitude which should be
approximately $1000$ times greater than Newtonian attraction. But
pressure of gravitons scattered by bodies gives a repulsion force
of the same order; the last is almost exact compensating this
attraction. To get Newton's law of gravity, it is necessary to
assume that gravitons form correlated pairs. By collision with a
body, such a pair should destruct in single gravitons. Flying away
gravitons of a pair should happen in independent directions, that
decreases a full cross-section of interaction with scattered
gravitons. As a result, an attraction force will exceed a
corresponding repulsion force acting between bodies. In such the
model, Newton's constant is connected with the Hubble constant
that gives a possibility to obtain a theoretical estimate of the
last. We deal here with a flat non-expanding universe fulfilled
with superstrong interacting gravitons; it changes the meaning of
the Hubble constant which describes magnitude of three small
effects of quantum gravity but not any expansion.
\par
The considered fine quantum mechanism of classical gravity differs
from a generally admitted one. In the full analogy with quantum
electrodynamics, it had been shown already in the first works in
quantum gravity \cite{5,6} that Newton's law may be explained as a
result of exchange with virtual longitudinal gravitons, sources of
which are attracting bodies.
\par
A conjecture about a composite nature of gravitons was considered
by few authors with other reasons (see short remarks and further
references in \cite{7}, and also the papers \cite{7a,7b}). The
main idea of works \cite{8,9}), where composite gravitons were
considered as correlated pairs of photons, seems to be the most
interesting for the author.

\section[2]{Screening the graviton background}

In author's papers \cite{1,2}, a cross-section $\sigma
(E,\epsilon)$ of interaction of a graviton with an energy
$\epsilon$ with any body having an energy $E$ was accepted to be
equal to:
\begin{equation}
\sigma (E,\epsilon)= D \cdot E \cdot \epsilon,
\end{equation}
where $D$ is some new dimensional constant. The Hubble
constant $H$ should be proportional to $D:$
\begin{equation}
H= {1 \over 2\pi} D \cdot \bar \epsilon \cdot (\sigma T^{4}),
\end{equation}
where $\bar \epsilon$ is an average graviton energy, $\sigma$ is
the Stephan-Boltzmann constant, $T$ is an effective temperature of
the graviton background. The interaction should be super-strong to
cause the whole redshift magnitude - it is necessary to have $D
\sim 10^{-27} m^{2}/eV^{2}.$
\par
If gravitons of the background run against a pair of bodies with
masses $m_{1}$ and $m_{2}$ (and energies $E_{1}$ and $E_{2}$) from
infinity, then a part of gravitons is screened. Let $\sigma
(E_{1},\epsilon)$ is a cross-section of interaction of body $1$
with a graviton with an energy $\epsilon=\hbar \omega,$ where
$\omega$ is a graviton frequency, $\sigma (E_{2},\epsilon)$ is the
same cross-section for body $2.$ In absence of body $2,$ a whole
modulus of a gravitonic pressure force acting on body $1$ would be
equal to:
\begin{equation}
4\sigma (E_{1},<\epsilon>)\cdot {1 \over 3} \cdot {4 f(\omega, T)
\over c},
\end{equation}
where $f(\omega, T)$ is a graviton spectrum with a temperature $T$
(assuming to be planckian), the factor $4$ in front of $\sigma
(E_{1},<\epsilon>)$ is introduced to allow all possible directions
of graviton running, $<\epsilon>$ is another average energy of
running gravitons with a frequency $\omega$ taking into account a
probability of that in a realization of flat wave a number of
gravitons may be equal to zero, and that not all of gravitons ride
at a body.
\par
Body $2,$ placed on a distance $r$ from body $1,$ will screen a
portion of running against body $1$ gravitons which is equal for
big distances between the bodies (i.e. by $\sigma
(E_{2},<\epsilon>) \ll 4 \pi r^{2}$):
\begin{equation}
\sigma (E_{2},<\epsilon>) \over 4 \pi r^{2}.
\end{equation}
Taking into account
all frequencies $\omega,$ an attractive force will act
between bodies $1$ and $2:$
\begin{equation}
F_{1}= \int_{0}^{\infty} {\sigma (E_{2},<\epsilon>) \over 4 \pi
r^{2}} \cdot 4 \sigma (E_{1},<\epsilon>)\cdot {1 \over 3} \cdot {4
f(\omega, T) \over c} d\omega.
\end{equation}
Let $f(\omega, T)$ is described with the Planck formula:
\begin{equation}
f(\omega,T)={{\omega}^{2} \over {4{\pi}^{2} c^{2}}}
{{\hbar \omega} \over {\exp(\hbar \omega/kT) - 1}}.
\end{equation}
Let $x \equiv {\hbar \omega/  kT},$ and $\bar{n} \equiv {1/
(\exp(x)-1)}$ is an average number of gravitons in a flat wave
with a frequency $\omega$ (on one mode of two distinguishing with
a projection of particle spin). Let $P(n,x)$ is a probability of
that in a realization of flat wave a number of gravitons is equal
to $n,$ for example $P(0,x)=\exp(-\bar{n}).$
\par
A quantity $<\epsilon>$ must contain the factor $(1-P(0,x)),$ i.e.
it should be:
\begin{equation}
<\epsilon> \sim \hbar \omega (1-P(0,x)),
\end{equation}
that let us to reject flat wave realizations with zero number of
gravitons.
\par
But attempting to define other factors in $<\epsilon>,$ we find
the difficult place in our reasoning. On this stage, it is
necessary to introduce some new assumption to find the factors.
Perhaps, this assumption will be well-founded in a future theory -
or would be rejected. If a flat wave realization, running against
a finite size body from infinity, contains one graviton, then one
cannot consider that it must stringent ride at a body to interact
with some probability with the one. It would break the uncertainty
principle by W. Heisenberg. We should admit that we know a
graviton trajectory. The same is pertaining to gravitons scattered
by one of bodies by big distances between bodies. What is a
probability that a single graviton will ride namely at the body?
If one denotes this probability as $P_{1},$ then for a wave with
$n$ gravitons their chances to ride at the body must be equal to
$n \cdot P_{1}.$ Taking into account the probabilities of values
of $n$ for the Poisson flux of events, an additional factor in
$<\epsilon>$ should be equal to $\bar{n} \cdot P_{1}.$ I admit
here that
\begin{equation}
P_{1}=P(1,x),
\end{equation}
where $P(1,x)=\bar{n}\exp(-\bar{n});$ (below it is admitted for
pairing gravitons: $P_{1}=P(1,2x)$ - see section 4).
\par
In such the case, we have for $<\epsilon>$ the following
expression:
\begin{equation}
<\epsilon>= \hbar \omega (1-P(0,x))\bar{n}^{2}\exp(-\bar{n}).
\end{equation}
Then we get for an attraction force $F_{1}:$
\begin{equation}
F_{1}= {4 \over 3}  {{D^{2} E_{1} E_{2}} \over {\pi r^{2} c}}
\int_{0}^{\infty} {{{\hbar}^{3} \omega^{5}} \over {4\pi^{2}c^{2}}}
(1-P(0,x))^{2}\bar{n}^{5}\exp(-2\bar{n}) d\omega =
\end{equation}
$${1 \over 3} \cdot {{D^{2} c (kT)^{6} m_{1} m_{2}} \over
{\pi^{3}\hbar^{3}r^{2}}} \cdot I_{1},$$
where
\begin{equation}
I_{1} \equiv \int_{0}^{\infty} x^{5}
(1-\exp(-(\exp(x)-1)^{-1}))^{2}(\exp(x)-1)^{-5}
\exp(-2(\exp(x)-1)^{-1}) dx=
\end{equation}
$$5.636 \cdot 10^{-3}.$$

If $F_{1}\equiv G_{1} \cdot  m_{1}m_{2}/r^{2},$ then the constant
$G_{1}$ is equal to:
\begin{equation}
G_{1} \equiv {1 \over 3} \cdot {D^{2} c(kT)^{6} \over
{\pi^{3}\hbar^{3}}} \cdot I_{1}.
\end{equation}
By $T=2.7 K:$
\begin{equation}
G_{1} =1215.4 \cdot G,
\end{equation}
that is three order greater than Newton's constant $G.$
\par
But if gravitons are elastic scattered with body $1,$ then our
reasoning may be reversed: the same portion (4) of scattered
gravitons will create a repulsive force $F_{1}^{'}$ acting on body
$2$ and equal to
\begin{equation}
F_{1}^{'} =F_{1},
\end{equation}
if one neglects with small allowances which are proportional to
$D^{3}/  r^{4}.$ The last ones are caused by decreasing of
gravitonic flux running against body $1$ due to screening by body
$2$ (see section 5).
\par
So, for bodies which elastic scatter gravitons, screening a flux
of single gravitons does not ensure Newtonian attraction. But for
gravitonic black holes which absorb any particles and do not
re-emit them (by the meaning of a concept, the ones are usual
black holes; I introduce a redundant adjective only from a
caution), we will have $F_{1}^{'} =0.$ It means that such the
object would attract other bodies with a force which is
proportional to $G_{1}$ but not to $G,$ i.e. Einstein's
equivalence principle would be violated for it. This conclusion,
as we shall see below, stays in force for the case of graviton
pairing too. The conclusion cannot be changed with taking into
account of Hawking's  quantum effect of evaporation of black holes
\cite{10}.
\section[3]{Graviton pairing}
To ensure an attractive force which is not equal to a repulsive
one, particle correlations should differ for {\it in} and {\it
out} flux. For example, single gravitons of running flux may
associate in pairs. If such pairs are destructed by collision with a
body, then quantities $<\epsilon>$ will distinguish for running
and scattered particles. Graviton pairing may be caused with
graviton's
own gravitational attraction or gravitonic spin-spin interaction.
Left an analysis of the nature of graviton pairing for the future;
let us see that gives such the pairing.
\par
To find an average number of pairs $\bar{n}_{2}$ in a wave with a
frequency $\omega$ for the state of thermodynamic equilibrium, one
may replace $\hbar \rightarrow 2\hbar$ by deducing the Planck
formula. Then an average number of pairs will be equal to:
\begin{equation}
\bar{n}_{2} ={1 \over {\exp(2x)-1}},
\end{equation}
and an energy of one pair will be equal to $2\hbar \omega.$ It is
important that graviton pairing does not change a number of
stationary waves, so as pairs nucleate from existing gravitons.
The question arises: how many different modes, i.e. spin
projections, may have graviton pairs? We consider that the
background of initial gravitons consists two modes. For massless
transverse bosons, it takes place as by spin $1$ as by spin $2.$
If graviton pairs have maximum spin $2,$ then single gravitons
should have spin $1.$ But from such particles one may constitute
four combinations: $\uparrow \uparrow, \ \downarrow \downarrow $
(with total spin $2$), and $\uparrow \downarrow, \
\downarrow\uparrow$ (with total spin $0).$ All these four
combinations will be equiprobable if spin projections $\uparrow$
and $\downarrow$ are equiprobable in a flat wave (without taking
into account a probable spin-spin interaction).
\par
But it is happened that, if expression (15)  is true, it follows
from the energy conservation law that composite gravitons should
be distributed only in two modes. So as
\begin{equation}
\lim_{x \to 0} {\bar{n}_{2} \over \bar{n}} ={1/2},
\end{equation}
then by $x \rightarrow 0$ we have $2\bar{n}_{2}=\bar{n},$ i.e. all
of gravitons are pairing by low frequencies. An average energy on
every mode of pairing gravitons is equal to $2 \hbar \omega
\bar{n}_{2},$ the one on every mode of single gravitons - $\hbar
\omega \bar{n}.$ These energies are equal by $x \rightarrow 0,$
because of it, the numbers of modes are equal too, if the
background is in thermodynamic equilibrium with surrounding
bodies.
\par
The above reasoning does not allow to choose a spin value $2$ or
$0$ for composite gravitons. A choice of namely spin $2$ would
ensure the following proposition: all of gravitons in one
realization of flat wave have the same spin projections. From
another side, a spin-spin interaction would cause it.
\par
The spectrum of composite gravitons is proportional to the Planck
one; it has the view:
\begin{equation}
f_{2}(2\omega,T)d\omega={\omega^{2} \over {4\pi^{2}c^{2}}} \cdot
{2\hbar \omega \over {\exp(2x)-1}}d\omega \equiv {(2\omega)^{2}
\over {32\pi^{2}c^{2}}} \cdot {2\hbar \omega \over
{\exp(2x)-1}}d(2\omega).
\end{equation}
It means that an absolute luminosity for the sub-system of
composite gravitons is equal to:
\begin{equation}
\int_{0}^{\infty} f_{2}(2\omega,T)d(2\omega)= {1 \over 8}\sigma
T^{4},
\end{equation}
where $\sigma$ is the Stephan-Boltzmann constant; i.e. an
equivalent temperature of this sub-system is
\begin{equation}
T_{2} \equiv (1/8)^{1/4}T = {2^{1/4} \over 2}T = 0.5946T.
\end{equation}
It is important that the graviton pairing effect does not changes
computed values of the Hubble constant and of anomalous
deceleration of massive bodies \cite{1}: twice decreasing of a
sub-system particle number due to the pairing effect is
compensated with twice increasing the cross-section of interaction
of a photon or any body with such the composite gravitons.
Non-pairing gravitons with spin $1$ give also its contribution in
values of redshifts, an additional relaxation of light intensity
due to non-forehead collisions with gravitons, and  anomalous
deceleration of massive bodies moving relative to the background
\cite{1,2}.

\section[4]{Computation of Newton's constant}
If running graviton pairs ensure for two bodies an attractive
force $F_{2},$ then a repulsive force due to re-emission of
gravitons of a pair alone will be equal to $F_{2}^{'} =F_{2}/2.$
It follows from that the cross-section for {\it single additional
scattered} gravitons of destructed pairs will be twice smaller
than for pairs themselves (the leading factor $2\hbar \omega$ for
pairs should be replaced with $\hbar \omega$ for single
gravitons). For pairs, we introduce here the cross-section $
\sigma (E_{2},<\epsilon_{2}>),$ where $<\epsilon_{2}>$ is an
average pair energy with taking into account a probability of that
in a realization of flat wave a number of graviton pairs may be
equal to zero, and that not all of graviton pairs ride at a body
($<\epsilon_{2}>$ is an analog of $<\epsilon>$).  This equality is
true in neglecting with small allowances which are proportional to
$D^{3}/ r^{4}$ (see section 5). Replacing $\bar{n} \rightarrow
\bar{n}_{2}, \hbar \omega \rightarrow 2\hbar \omega,$ and $P(n,x)
\rightarrow P(n,2x),$ where $P(0,2x)= \exp(-\bar{n}_{2}),$ we get
for graviton pairs:
\begin{equation}
<\epsilon_{2}> \sim 2\hbar \omega
(1-P(0,2x))\bar{n}_{2}^{2}\exp(-\bar{n}_{2}).
\end{equation}
This expression does not take into account only that beside pairs
there may be single gravitons in a realization of flat wave. To
reject cases when,instead of a pair, a single graviton runs
against a body (a contribution of such gravitons in attraction and
repulsion is the same), we add the factor $P(0,x)$ into
$<\epsilon_{2}>:$
\begin{equation}
<\epsilon_{2}> = 2\hbar \omega
(1-P(0,2x))\bar{n}_{2}^{2}\exp(-\bar{n}_{2}) \cdot P(0,x).
\end{equation}
Then a force of attraction of two bodies due to pressure of
graviton pairs $F_{2}$ - in the full analogy with (5) - will be
equal to \footnote{In initial version of this paper, factor 2 was
lost in the right part of Eq. (22), and the theoretical values of
$D$ and $H$ were overestimated of $\sqrt{2}$ times}:
\begin{equation}
F_{2}= \int_{0}^{\infty} {\sigma (E_{2},<\epsilon_{2}>) \over 4
\pi r^{2}} \cdot 4 \sigma (E_{1},<\epsilon_{2}>)\cdot {1 \over 3}
\cdot {4 f_{2}(2\omega,T) \over c} d\omega =
\end{equation}
$$ {8 \over 3} \cdot
{D^{2} c(kT)^{6} m_{1}m_{2} \over {\pi^{3}\hbar^{3}r^{2}}}\cdot
I_{2},$$
where
\begin{equation}
I_{2} \equiv \int_{0}^{\infty}{ x^{5}
(1-\exp(-(\exp(2x)-1)^{-1}))^{2}(\exp(2x)-1)^{-5} \over
\exp(2(\exp(2x)-1)^{-1}) \exp(2(\exp(x)-1)^{-1})} d x =
\end{equation}
$$2.3184 \cdot 10^{-6}.$$
The difference $F$ between attractive and repulsive forces will be
equal to:
\begin{equation}
F \equiv F_{2}- F_{2}^{'}={1 \over 2}F_{2} \equiv G_{2}{m_{1}m_{2}
\over r^{2}},
\end{equation}
where the constant $G_{2}$ is equal to:
\begin{equation}
G_{2} \equiv {4 \over 3} \cdot {D^{2} c(kT)^{6} \over
{\pi^{3}\hbar^{3}}} \cdot I_{2}.
\end{equation}
As $G_{1}$ as well $G_{2}$ are proportional to $T^{6}$ (and $H
\sim T^{5},$ so as $\bar{\epsilon} \sim T$).
\par
If one assumes that $G_{2}=G,$ then it follows from (25) that by
$T=2.7K$ the constant $D$ should have the value:
\begin{equation}
D=0.795 \cdot 10^{-27}{m^{2} / eV^{2}}.
\end{equation}
An average graviton energy of the background is equal to:
\begin{equation}
\bar{\epsilon} \equiv \int_{0}^{\infty} \hbar \omega \cdot
{f(\omega, T) \over \sigma T^{4}} d \omega = {15 \over
\pi^{4}}I_{4}kT,
\end{equation}
where
$$I_{4} \equiv \int_{0}^{\infty} {x^{4} dx \over
{\exp(x)-1}}=24.866 $$
(it is $\bar{\epsilon}=8.98 \cdot 10^{-4}eV$ by $T=2.7K$).
\par
We can use (2) and (25) to establish a connection between the two
fundamental constants $G$ and $H$ under the condition that
$G_{2}=G.$ We have for $D:$
\begin{equation}
D= {2\pi H \over \bar{\epsilon}
\sigma T^{4}}= {2 \pi^{5} H \over 15 k \sigma T^{5} I_{4}};
\end{equation}
then
\begin{equation}
G=G_{2} = {4 \over 3} \cdot {D^{2} c(kT)^{6} \over
{\pi^{3}\hbar^{3}}} \cdot I_{2}= \\
{64 \pi^{5} \over 45} \cdot {H^{2}c^{3}I_{2} \over \sigma T^{4}
I_{4}^{2}}.
\end{equation}
So as the value of $G$ is known much better than the value of $H,$
let us express $H$ via $G:$
\begin{equation}
H= (G  {45 \over 64 \pi^{5}}  {\sigma T^{4} I_{4}^{2} \over
{c^{3}I_{2}}})^{1/2}= 2.14 \cdot 10^{-18}s^{-1},
\end{equation}
or in the units which are more familiar for many of us: $H=66.875
\ km \cdot s^{-1} \cdot Mpc^{-1}.$
\par
This value of $H$ is is in the good accordance with the majority
of present astrophysical estimations \cite{3,12}, but it is lesser
than some of them \cite{12a} and than it follows from the observed
value of anomalous acceleration of Pioneer 10 \cite{4} $w=(8.4 \pm
1.33)\cdot 10^{-10} \ m/s^{2}.$ Any massive body, moving relative
to the background, must feel a deceleration $w \simeq Hc$
\cite{1,2}; with $H=2.14 \cdot 10^{-18}s^{-1}$ we have $Hc=6.419
\cdot 10^{-10} \ m/s^{2}.$
\par
The observed value of anomalous acceleration of Pioneer 10 should
represent the vector difference of the two acceleration: an
acceleration of Pioneer 10 relative to the graviton background,
and  an acceleration of the Earth relative to the background.
Possibly, the last is displayed as an annual periodic term in the
residuals of Pioneer 10 \cite{11}. If the solar system moves with
a noticeable velocity relative to the background, the Earth's
anomalous acceleration projection on the direction of this
velocity will be smaller than for the Sun - because of the Earth's
orbital motion. It means that in a frame of reference, connected
with the Sun, the Earth should move with an anomalous acceleration
having non-zero projections as well on the orbital velocity
direction as on
the direction of solar system motion relative to the background.
Under some conditions, the Earth's anomalous acceleration in this
frame of reference may be periodic. The axis of Earth's orbit
should feel an annual precession by it.

\section[5]{Why and when gravity is geometry}
The described quantum mechanism of classical gravity gives
Newton's law with the constant $G_{2}$ value (25) and the
connection (29) for the constants $G_{2}$ and $H.$  We have
obtained the rational value of $H$ (30) by $G_{2} = G,$ if the
condition of big distances is fulfilled:
\begin{equation}
\sigma (E_{2},<\epsilon>) \ll 4 \pi r^{2}.
\end{equation}
Because it is known from experience that for big bodies of the
solar system, Newton's law is a very good approximation, one would
expect that the condition (30) is fulfilled, for example, for the
pair Sun-Earth. But assuming $r=1 \ AU$ and
$E_{2}=m_{\odot}c^{2},$ we obtain assuming for rough estimation
$<\epsilon> \rightarrow
\bar{\epsilon}:$ $${\sigma (E_{2},<\epsilon>) \over 4 \pi r^{2}}
\sim 4 \cdot 10^{12}. $$ It means that in the case of interaction
of gravitons or graviton pairs with the Sun in the aggregate, the
considered quantum mechanism of classical gravity could not lead
to Newton's law as a good approximation. This "contradiction" with
experience is eliminated if one assumes that gravitons interact
with "small particles" of matter - for example, with atoms. If the
Sun contains of $N$ atoms, then $\sigma (E_{2},<\epsilon>)=N \sigma
(E_{a},<\epsilon>),$ where $E_{a}$ is an average energy of one
atom. For rough estimation we assume here that $E_{a}=E_{p},$ where
$E_{p}$ is a proton rest energy; then it is $N \sim 10^{57},$ i.e.
${\sigma (E_{a},<\epsilon>)/ 4 \pi r^{2}} \sim  10^{-45} \ll 1.$
\par
This necessity of "atomic structure" of matter for working the
described quantum mechanism is natural relative to usual
bodies. But would one expect that black holes have a similar
structure? If any radiation cannot be emitted with a black hole, a
black hole should interact with gravitons as an aggregated object,
i.e. the condition (31) for a black hole of sun mass has not been
fulfilled even at distances $\sim 10^{6} \ AU.$
\par
For bodies without an atomic structure, the allowances, which are
proportional to $D^{3}/ r^{4}$ and are caused by decreasing a
gravitonic flux due to the screening effect, will have a factor
$m_{1}^{2}m_{2}$ or $m_{1}m_{2}^{2}.$ These allowances break the
equivalence principle for such the bodies.
\par
For bodies with an atomic structure, a force of interaction is
added up from small forces of interaction of their "atoms": $$
F \sim N_{1}N_{2}m_{a}^{2}/r^{2}=m_{1}m_{2}/r^{2},$$ where
$N_{1}$ and $N_{2}$ are numbers of atoms for bodies $1$ and $2$.
The allowances to full forces due to the screening effect will be
proportional to the quantity: $N_{1}N_{2}m_{a}^{3}/r^{4},$ which
can be expressed via the full masses of bodies as
$m_{1}^{2}m_{2}/r^{4}N_{1}$ or $m_{1}m_{2}^{2}/r^{4}N_{2}.$ By big
numbers $N_{1}$ and $N_{2}$ the allowances will be small. The
allowance to the force $F,$ acting on body $2,$ will be equal to:
\begin{equation}
\Delta F ={1 \over 2 N_{2}} \int_{0}^{\infty} {\sigma^{2}
(E_{2},<\epsilon_{2}>) \over (4 \pi r^{2})^{2}} \cdot 4 \sigma
(E_{1},<\epsilon_{2}>)\cdot {1 \over 3} \cdot {4 f_{2}(2\omega,T)
\over c} d\omega =
\end{equation}
$${2 \over 3N_{2}} \cdot {{D^{3} c^{3} (kT)^{7}
m_{1} m_{2}^{2}} \over {\pi^{4}\hbar^{3}r^{4}}} \cdot I_{3},$$
(for body $1$ we shall have the similar expression if replace
$N_{2} \rightarrow N_{1}, \ m_{1}m_{2}^{2} \rightarrow
m_{1}^{2}m_{2}$), where
$$ I_{3} \equiv \int_{0}^{\infty} {x^{6}
(1-\exp(-(\exp(2x)-1)^{-1}))^{3}(\exp(2x)-1)^{-7} \over
\exp(3(\exp(x)-1)^{-1})} d x = 1.0988 \cdot 10^{-7}. $$
\par
Let us find the ratio:
\begin{equation}
{\Delta F \over F} = {D E_{2} kT \over {N_{2} 2\pi r^{2}}} \cdot
{I_{3} \over I_{2}}.
\end{equation}
Using this formula, we can find by $E_{2}=E_{\odot}, \ r=1 \ AU:$
\begin{equation}
{\Delta F \over F} \sim 10^{-46}.
\end{equation}
\par
An analogical allowance to the force $F_{1}$ has by the same
conditions the order $\sim 10^{-48}F_{1},$ or $\sim 10^{-45}F.$
One can replace $E_{p}$ with a rest energy of very big atom - the
geometrical approach will left a very good language to describe
the solar system. We see that for bodies with an atomic structure
the considered mechanism leads to very small deviations from
Einstein's equivalence principle, if the condition (31) is
fulfilled for microparticles, which prompt interact with
gravitons.
\par
For small distances we shall have:
\begin{equation}
\sigma (E_{2},<\epsilon>) \sim 4 \pi r^{2}.
\end{equation}
It takes place by $E_{a}=E_{p}, \ <\epsilon> \sim 10^{-3} \ eV$
for $r \sim 10^{-11} \ m.$ This quantity is many order larger than
the Planck length. The equivalence principle should be broken at
such distances.
\par
Under the condition (35), big digressions from Newton's law will
be caused with two factors: 1) a screening portion of a running
flux of gravitons is not small and it should be taken into account
by computation of the repulsive force; 2) a value of this portion
cannot be defined by the expression (4).
\par
Instead of (4), one might describe this portion at small distances
with an expression of the kind:
\begin{equation}
{1 \over 2}(1+ {\sigma
(E_{a},<\epsilon>)/ \pi r^{2}}-(1+ {\sigma (E_{a},<\epsilon>)/ \pi
r^{2}})^{1/2} )
\end{equation}
(the formula for a spheric segment area is used
here \cite{13}). Formally, by ${\sigma (E_{a},<\epsilon>)/ \pi
r^{2}} \rightarrow \infty$ we shall have for the portion (36):
$$\sim {1 \over 2}({\sigma (E_{a},<\epsilon>)/ \pi r^{2}}-({\sigma
(E_{a},<\epsilon>)/ \pi})^{1/2}/r),$$ where the second term shows
that the interaction should be weaker at small distances.  We
might expect that a screening portion may tend to a fixing value
at super-short distances. But, of course, at such distances the
interaction will be super-strong and our naive approach would be
not valid.

\section[6]{Conclusion}
It is known that giant intellectual efforts to construct a quantum
theory of metric field, based on the theory of general relativity,
have not a hit until today (see the recent review \cite{18}). From
a point of view of the considered approach, one may explain it by
the fact that gravity is not geometry at short distances $\sim
10^{-11} \ m.$ Actually, it means that at such the distances quantum
gravity cannot be described alone but only in some unified manner,
together with other interactions including the strong one.
\par
It follows from section 5 of the present work that the geometrical
description of gravity should be a good idealization at big distances
by the
condition of "atomic structure" of matter. This condition cannot
be accepted only for black holes which must interact with
gravitons as aggregated objects. In addition, the equivalence
principle is roughly broken for black holes, if the described
quantum mechanism of classical gravity is realized in the nature.
\par
Other important features of this mechanism are the following ones.
$\\ \bullet$ Attracting bodies are not initial sources of
gravitons. In this sense, a future theory must be non-local to
describe gravitons running from infinity. Non-local models were
considered by Efimov in his book \cite{14}. The idea to describe
gravity as an effect caused by running ab extra particles was
criticized by the great physicist Richard Feynman in his public
lectures at Cornell University \cite{15}, but the Pioneer 10
anomaly \cite{4}, perhaps, is a good contra argument pro this
idea. \\ $\bullet$ Newton's law takes place if gravitons are
pairing; to get preponderance of attraction under repulsion,
graviton pairs should be destructed by interaction with matter
particles. \\ $\bullet$ The described quantum mechanism of
classical gravity is obviously asymmetric relative to the time
inversion. By the time inversion, single gravitons would run
against bodies forming pairs. It would lead to replacing a body
attraction with a repulsion. But such the change will do
impossible graviton pairing. Cosmological models with the
inversion of the time arrow were considered by Sakharov \cite{16}.
Penrose reasoned about a hidden physical law determining the time
arrow direction \cite{17}; it will be interesting if realization
in the nature of Newton's law determines this direction.
\\ $\bullet$ The two fundamental constants - Newton's and Hubble's
ones - are connected with each other in such the model. The
estimate of Hubble's constant has been got here using an
additional postulate $P_{1}=P(1,2x)$ for pairing gravitons.
\\ $\bullet$ It is proven that graviton pairs should be
distributed in two modes with different spin projections.
\\ $\bullet$ From thermodynamic reasons, it is assumed here that
the graviton background has the same temperature as the microwave
background. Also it follows from the condition of detail
equilibrium, that both backgrounds should have the planckian
spectra. Composite gravitons will have spin $2$, if single
gravitons have the same spin as photons. The question arise, of
course: how are gravitons and photons connected? Has the
conjecture by Adler et al. \cite{8} (that a graviton with spin $2$
is composed with two photons) chances to be true? Intuitive demur
calls forth a huge self-action, photons should be endued with
which - but one may get a unified theory on this way. To verify
this conjecture in experiment, one would search for transitions in
interstellar gas molecules caused by the microwave background,
with an angular momentum change corresponding to absorption of
spin $2$ particles (photon pairs). A frequency of such the
transitions should correspond to an equivalent temperature  of the
sub-system of these composite particles $T_{2}=0.5946 T,$ if $T$
is a temperature  of the microwave background.
\par
A future theory dealing with gravitons as usual particles having
an energy, a momentum etc ("gravitonics" would be a fine name for
it) should have a number of features, which are not characterizing
any existing model, to image the recounted above features of a
possible quantum mechanism of gravity.
\par
The main results of this work were poster presented at MG10 and
Thinking'03 \cite{28,29}.


\end{document}